\documentclass[aip,jap,reprint,floatfix,groupedaddress]{revtex4-1}
\usepackage{graphicx}
\usepackage{hyperref}
\usepackage{amsmath,amssymb,amsfonts}

\DeclareMathOperator{\Var}{Var}

\begin{document}

\title[Electrical conductance of two-dimensional composites with embedded rodlike fillers]{Electrical conductance of two-dimensional composites with embedded rodlike fillers: an analytical consideration and comparison of two computational approaches}

\author{Yuri~Yu.~Tarasevich}
\email[Corresponding author: ]{tarasevich@asu.edu.ru}

\author{Irina~V.~Vodolazskaya}
\email{vodolazskaya\_agu@mail.ru}

\author{Andrei~V.~Eserkepov}
\email{dantealigjery49@gmail.com}

\author{Renat~K.~Akhunzhanov}
\email{akhunzha@mail.ru}

\affiliation{Laboratory of Mathematical Modeling, Astrakhan State University, Astrakhan, Russia, 414056}

\date{\today}

\begin{abstract}
Using Monte Carlo simulation, we studied the electrical conductance of two-dimensional films. The films consisted of a poorly conductive host matrix and highly conductive rodlike fillers (rods). The rods were of various lengths, obeying  a log-normal distribution. They were allowed to be aligned along a given direction. The impacts of length dispersity and the extent of rod alignment on the insulator-to-conductor phase transition were studied.
Two alternative computational approaches were compared. Within Model~I, the films were transformed into resistor networks with regular structures and randomly distributed conductances. Within Model~II, the films were transformed into resistor networks with irregular structures but with equal conductivities of the conductors. Comparison of the models  evidenced similar behavior in both models when the concentration of fillers exceeded the percolation threshold. Some analytical results were obtained: (i)~the relationship between the number of fillers per unit area and the transmittance of the film within Model~I, (ii)~the electrical conductance of the film for dense networks within Model~II.
\end{abstract}


\maketitle

\section{Introduction}\label{sec:intro}
Thin films with elongated conductive fillers having high aspect ratios (the ratio of the characteristic length to the characteristic transverse dimension of each particle) can exhibit both high electrical conductance and  excellent optical transparency.\cite{Forro2018ACSN} These features have led to a widespread use of such composites in transparent conductors (TCs), transistors, sensors, and flexible and stretchable conductors.\cite{Hecht2011AM,Forro2018ACSN} Nowadays, some two-dimensional (2D) systems such as transparent electrodes present examples of where highly conductive particles, e.g., nanowires (NWs),  nanotubes (NTs), and nanorods (NRs), form a random resistor network (RRN) inside a poorly  conductive host matrix (substrate).\cite{Mutiso2015PPS,Kumar2016JAP,Kumar2017JAP} The appearance of a percolation cluster in similar systems drastically changes their physical properties and is associated with an insulator-to-conductor phase transition. Length dispersity is common for NWs, NTs, and NRs.\cite{Kyrylyuk2008PNAS,Gnanasekaran2014,Borchert2015N,Majidian2017SR} These works evidenced that the length distributions of NWs, NTs, and NRs are close to log-normal distributions. Furthermore, alignment of such elongated objects may be produced in different ways.\cite{Chen2004JVSTB,Du2005PRB,Park2006JPSB,Behnam2007JAP,Yang2011NML} Both length dispersity and alignment affect the electrical conductance of the samples.\cite{Du2005PRB,Hicks2009PRE,Jagota2015SR,Ackermann2016SR} The aspect ratios of metallic NWs or carbon NTs (CNTs) are of the order of hundreds and even thousands.\cite{Khanarian2013JAP,Borchert2015N,Forro2013AIPA}

Recently, a theoretical model for the effect of nanowire dimensions and coverage on the coupled relationship between sheet resistance and transmission in conductive nanowire networks has been presented.~\cite{Ainsworth2018} The model assumes the transmission of light through the network to be proportional to its fractional open area. It predicts that the sheet resistance is proportional to the square of the nanowire diameter and inversely proportional to the squares of both the nanowire length and the network coverage.

The study of percolation in 2D systems of rodlike particles or sticks and its connection with electrical conductance has a long history.\cite{Pike1974PRB,Balberg1982SSC,Balberg1983PRL} Some geometrical properties of 2D systems of sticks or fibers have been derived.\cite{Goudsmit1945RMP,Corte1969,Yi2004JAP,Heitz2004NT,Callaghan2016PCCP,Kim2018JAP} Numerous works have been devoted to the electrical properties of RRNs produced by randomly deposited sticks\cite{Kim2018JAP} and different kinds of templates.\cite{Kumar2016JAP,Kumar2017JAP} The topology of connected carbon nanotubes in random networks has also been investigated.\cite{Heitz2004NT} Analytical expressions for some useful quantities, e.g., mean contact probability, mean contact angle and number of contacts per CNT, have been derived and tested by means of Monte Carlo simulations. Similar results have also been reported recently.\cite{Callaghan2016PCCP}

Recently, an analytical formula for the sheet resistance of the dense homogeneous and isotropic RRNs has been proposed from geometrical considerations.\cite{Kumar2016JAP,Kumar2017JAP} The main idea of the consideration is that, in a dense homogeneous and isotropic 2D RRN, the electrical potential is expected to change linearly between two electrodes applied to the opposite borders of such an RRN. This assumption is supported by both extensive computer simulations\cite{Forro2018ACSN} and experiments.\cite{Sannicolo2018ACSN} Thus, even slightly above the percolation threshold $n = 1.23 n_c$ where $n_c$ is the critical (percolating) number density, i.e., the number of fillers per unit area, the potential decrease is fitted by a linear function with the coefficient of determination $R^2 = 0.996$ while for denser systems, $n = 2 n_c$, $R^2 = 0.998$.\cite{Forro2018ACSN}

Two approaches are frequently used to calculate the electrical conductance of 2D composites. In the first one, a continuous composite is divided into cells and each cell is treated as a set of conductors. In such a way, the conductor is transformed into an RRN with a regular structure but randomly distributed resistivities.\cite{Yuge1978JPC} Both the conductivity of the host matrix and the conductivities of fillers are accounted for in this approach. It can also represent composites with concentrations of fillers both below and above the percolation threshold. Simulations of the electrical conductance of 2D systems with rodlike fillers of equal length have been performed using this method.\cite{Tarasevich2016PRE,Tarasevich2017PhysA,Tarasevich2018JPhCS,Tarasevich2018PREa,Tarasevich2018PREb}
We will denote this approach as Model~I. By contrast, within the second approach, the host matrix is completely ignored; the conducting fillers are considered to form an RRN of irregular geometry but with equal conductivity of its conductors.\cite{Behnam2007JAP,Hicks2009PRE,Mietta2014JCP,GomesdaRocha2015NS,Jagota2015SR,Jagota2015SR,Zezelj2016SST,Kumar2016JAP,Kumar2017JAP,Schiessel2017PRM,He2018JAP,Ni2018NT,Hicks2018JAP,Kim2018JAP} Obviously, only composites with filler concentrations exceeding the percolation threshold can be considered using  this method. We will denote this approach as Model~II. In some cases, the effects of junction resistances have additionally been taken into  account.\cite{Kang2011pssb,Zezelj2012PRB,Callaghan2016PCCP,Forro2018ACSN,Han2018SR} Recently, the equivalence between the sheet resistances of disordered networks and those of regular ordered networks has been established when the resistance of interwire contacts has been considered.\cite{Callaghan2016PCCP}

To the best of our knowledge, for the case of anisotropic RRNs composed of aligned rods with length dispersity, no comparison of the two approaches has been presented in the literature. In this research, we have tried to perform such a comparison. One can expect that, above the percolation threshold, the electrical conductance calculated within a discrete approach tends to the electrical conductance calculated using a continuous approach when the cell size tends to zero. Furthermore, using both approaches, we have examined the effect of dispersity of filler length and filler alignment on the electrical conductance and the transmittance of 2D composites with rodlike fillers.

The rest of the paper is constructed as follows. In Sec.~\ref{sec:methods}, the technical details of the simulations and calculations are described. Section~\ref{sec:results} presents our main findings. Section~\ref{sec:concl} summarizes the main results.

\section{Methods}\label{sec:methods}
\subsection{Sampling}
\begin{figure}[!htb]
  \centering
  \includegraphics[width=\columnwidth]{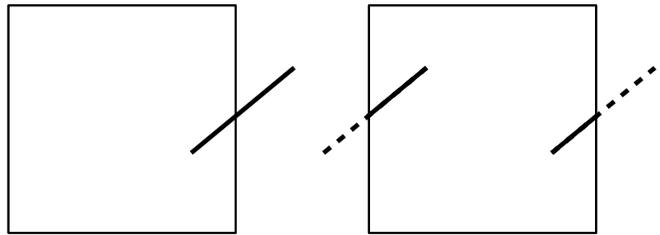}
  \caption{Application of PBCs to a rod, which intersects a border of the substrate.\cite{Lee2015JMS}\label{fig:replica}}
\end{figure}

Zero-width (widthless) rods were deposited randomly and uniformly with given anisotropy onto a substrate  of size $L \times L$ and with periodic boundary conditions (PBCs), i.e., onto a torus. Intersections of the rods were allowed. If a rod intersected a border of the substrate, a replica was created according to the PBCs, i.e., the rod was transformed into two shorter rods which touching the opposite borders of the substrate [Fig.~\ref{fig:replica}].

The lengths of the rods, $l$, varied according to a log-normal distribution with the probability density function (PDF)
\begin{equation}\label{eq:lognorm}
  f_l(l)=\frac{1}{l\sigma_l\sqrt{2\pi }}\exp \left( -\frac{\left( \ln l-\mu_l \right)^2}{2\sigma_l^2} \right).
\end{equation}
All our computations were performed for a fixed value of the mean length, viz., $\langle l \rangle = 1$.
For this particular value of the mean, the parameters of the log-normal distribution are
$$
\mu_l = -\frac{\sigma_l^2}{2}, \quad  \sigma_l^2 = \ln \left(\sigma (l)^2 + 1\right),
$$
where $\sigma (l)$ is the desired value of the standard deviation of the length dispersity.
In such a way, we could extract and study the individual effect of the length dispersity. All our simulations were performed for $L = 32 \langle l \rangle$.

The anisotropy of the system was characterized by  the order parameter (see, e.g., Ref.~\onlinecite{Frenkel1985PRA})
\begin{equation}\label{eq:s}
s=N^{-1}\sum_{i=1}^N \cos 2\theta_i = 2 \langle \cos^2 \theta \rangle  - 1,
\end{equation}
where $\theta_i$ is the angle between the axis of the $i$-th rod and the horizontal axis $x$, and $N$ is the total number of rods in the system.

In our simulations, the angles were distributed according to the normal distribution with a  PDF\cite{Tarasevich2018PREa}
\begin{equation*}\label{eq:angledistrib}
  f_\theta(\theta) = \frac{1}{\sqrt{- \pi \ln s}} \exp\left(\frac{\theta^2}{\ln s} \right),
\end{equation*}
where the variance is connected with the order parameter as
\begin{equation*}\label{eq:variance}
  \Var \theta = -0.5 \ln  s.
\end{equation*}

We performed our simulations for different values of the order parameter and length dispersity. For each sample, a sequence of random positions (two coordinates for each rod), orientations, and lengths was generated. This sequence was used to produce a film with the desired anisotropy, $s$, and number density of the rods, $n$,
\begin{equation*}\label{eq:numdens}
  n = \frac{N}{L^2}.
\end{equation*}
Since support of the log-normal distribution is $l \in (0, \infty)$, the probability that $l>L$ is finite, although very small. All rods with $l>L$ were rejected for deposition and excluded from the sequence.

\subsection{Computation of the electrical conductance}
In our study, the contact resistance between rods was ignored, i.e., the conductance of the system under consideration was completely defined by the conductivity of the rods themselves. For each sample, the electrical conductance was calculated along two mutually perpendicular directions.

\subsubsection{Discrete approach (Model~I)}
This approach (Model~I) is based on the idea described in Ref.~\onlinecite{Yuge1978JPC}. The rods were considered to be highly conductive whilst the substrate was assumed to be poorly conductive. In all our computations, the electrical contrast, i.e., the ratio of the conductivity of the rods, $\gamma_p$, to the conductivity of the substrate, $\gamma_m$, was $\Delta = 10^6$. To account the electrical conductivity of both the substrate and the fillers, a three-step transformation of the samples was performed.\cite{Lebovka2018PRE,Tarasevich2018PREa,Tarasevich2018PREb}
\paragraph{In the first step,} a sample was divided into square cells of equal size, in such the way, that the film was transformed into a mesh of $m \times m$  cells.

\paragraph{In the second step,} each cell of the mesh was marked as conductive and opaque when it contained any part of a rod or some parts of several rods, otherwise the cell was marked as insulating and transparent. In such a way, the continuous problem of rods was transformed into a problem of linear polyominoes.\cite{Lebovka2018PRE,Tarasevich2018PREa,Tarasevich2018PREb} The fraction of conductive cells was denoted as the concentration or the filling fraction, $p$. The quantity $T = 1 - p$ was treated as the transmittance of the film. For isotropically deposited rods of equal lengths ($s=0$, $l=1$, $\sigma(l) = 0$), the mean size of polyominoes is equal to
$$
\langle k \rangle = \frac{4 m}{\pi L} + 1.
$$
In the case of strictly aligned rods of equal length ($s=1$, $l=1$, $\sigma(l) = 0$), the polyominoes are simply rectangles with the aspect ratio\cite{Tarasevich2018PREa}
\begin{equation*}\label{eq:k}
\langle k \rangle  = \frac{m}{L} + 1.
\end{equation*}

\paragraph{In the third step,} the Hoshen--Kopelman algorithm\cite{HK76} was applied to check whether there were any percolating (spanning) clusters of conductive cells.  Then, each cell was replaced by four conductors connected as a four-pointed star (crosswise).\cite{Yuge1978JPC} All four conductors were assumed to have the same conductivities, viz., $\gamma_p = 2 \times 10^6 \times m/L$ arb. units when the cell was marked as conductive and $\gamma_m = 2 m/L$ arb. units otherwise. In such the way, the film was transformed into an RRN. The factor $m/L$ ensures the constant electrical conductance of a rod located parallel to a system border at any mesh size. The conductance of such an RRN was calculated using the Frank--Lobb  algorithm\cite{FL88} for each desired number density in both mutually perpendicular directions.  As a result, the electrical conductance versus the number density, $G_i(n)$, the electrical conductance against the  concentration, $G_i(p)$, and the presence of a percolation cluster depending on the number density and concentration were obtained for each sample. Here, $i=\parallel, \perp$ means the directions along and perpendicular to the alignment of the fillers, respectively. For each pair  of values $s$ and $\sigma(l)$, the electrical conductances were computed for the meshes $m = 256, 512, 1024, 2048$.
To characterize the anisotropy of the conductance, the ratio of the conductivities\cite{Tarasevich2016PRE,Tarasevich2017PhysA,Tarasevich2018JAP} was utilized
\begin{equation}\label{eq:delta}
  \delta = \frac{\left| \log_{10} G_\parallel / G_\perp \right|}{\log_{10} \Delta}.
\end{equation}

\subsubsection{Continuous approach (Model~II)}

\begin{figure}[!htb]
  \centering
  \includegraphics[width=\columnwidth]{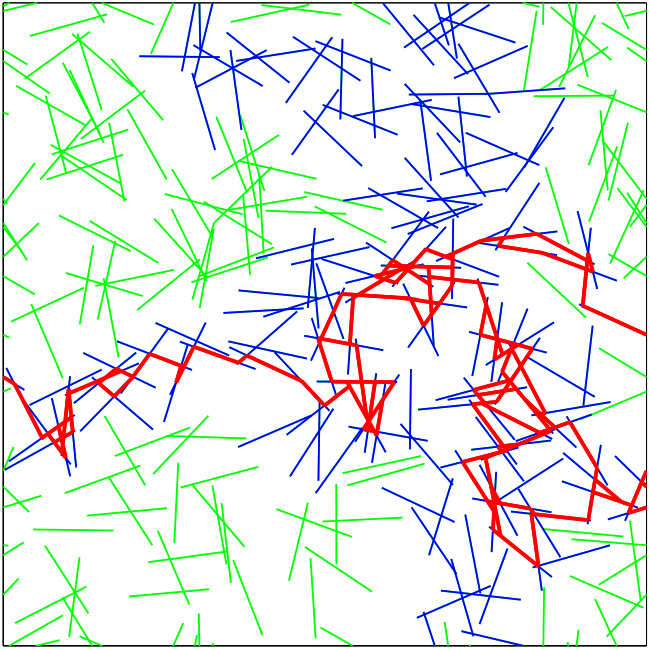}
  \caption{Example of a random graph produced by isotropic deposition of equal-length rods onto a substrate with linear size $L=8$. Incipient percolation cluster and its geometrical backbone are shown.\label{fig:clusters}}
\end{figure}

In the continuous approach (Model~II), rods were sequentially deposited onto a square substrate until a cluster appeared which spanned the opposite borders of the system. To detect a spanning cluster, the Union--Find algorithm\cite{Newman2000PRL,Newman2001PRE} modified for continuous systems\cite{Mertens2012PRE,Li2009PRE}  was applied. When a spanning cluster was found, all other clusters were removed. Then, the geometrical backbone was extracted using the Grassberger algorithm\cite{Grassberger1992JPhA} adapted to a continuous percolation of rods [Fig.~\ref{fig:clusters}]. The geometrical backbone is the set of all sites that are connected by their edges to both opposite borders of the system by two paths having no edges in common. Except for ``Wheatstone bridges'', it consists exactly of those sites and edges through which current would flow if the two opposite borders of the system were subject to a potential difference. The current-carrying part of the percolation cluster is called the electrical backbone. Thus, the electrical backbone is the geometrical backbone without the perfectly balanced edges.  [For a review of different algorithms intended for backbone identification, see Ref.~\onlinecite{Tarasevich2018JPhCSBB}.] At the next stage, an adjacency matrix was formed for the geometrical backbone. Having this adjacency matrix in hand, Kirchhoff's current law can be used for each junction of rods, and Ohm's law for each circuit between two junctions. The resulting set of equations was solved using MATLAB to find the total conductance of the RRN.

\section{Results}\label{sec:results}

\subsection{Relationship between the number density and the transmittance (Model~I)}

\begin{figure}[!htb]
  \centering
  \includegraphics[width=0.75\columnwidth]{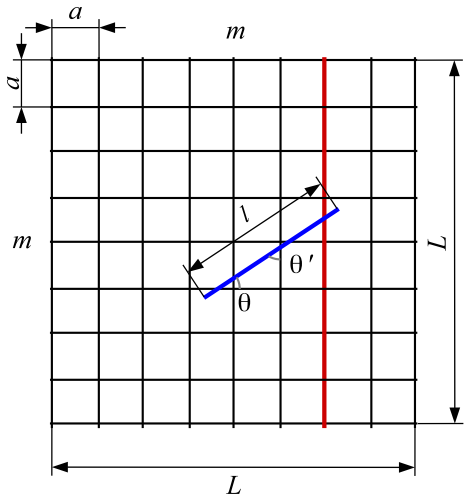}
  \caption{System under consideration. When the left end of the rods is located at a distance not exceeding $l \cos \theta$ from the vertical line, the intersection of the rod with the line is ensured.\label{fig:mesh}}
\end{figure}

Consider a domain of size $L \times L$ with PBCs. The domain is covered by a square mesh $m \times m$. Each square cell has a linear size $a = L/m$ [Fig.~\ref{fig:mesh}]. Let us randomly and homogeneously deposit rods onto  this domain. The rods may have length dispersity and obey an angular distribution. Let the length of the rods obey the PDF $f_l(l)$, while their angular distribution corresponds to the PDF  $f_\theta(\theta)$. However, thee maximal rod length, $l_m$, cannot exceed the domain size, i.e.,  $l_m < L$. We are looking for the fraction of empty (transparent) cells, i.e., the cells of the mesh that contain no part of any rod.

Since the deposition of a rod is independent of other rods, after the deposition of $N$ rods, the expected value of the number of empty cells is
\begin{equation*}\label{eq:M}
  M = m^2 (  1 - P )^N,
\end{equation*}
where $P = P_1 + P_2$ is the probability that a rod intersects any given cell or is located within it. Here, $P_1$ is the probability that a rod starts in the cell, and $P_2$ is the probability that a rod intersects that cell from outside.

Let us suppose that the position of the rod is given by the coordinates of one of its ends and by an angle, $\theta$, between the rod and one of the Cartesian axes ($-\frac{\pi}{2} \leqslant \theta < \frac{\pi}{2}$). Due to homogeneous deposition, the probability that the rod starts in the given cell is
\begin{equation*}\label{eq:P1}
  P_1 = \frac{a^2}{L^2} = m^{-2}.
\end{equation*}
This probability depends neither on the length dispersity of the rods nor on their angular distribution.

Consider a rod of length $l$ with orientation $\theta$ in relation to the abscissa. Such a rod intersects a vertical line if and only if the distance between its starting end and the line obeys the inequality $x \leqslant l \cos \theta$ [Fig.~\ref{fig:mesh}]. The probability that this inequality holds is
 $$
 \frac{ L l \cos \theta}{L^2} = \frac{ l \cos \theta}{L}.
 $$
The probability that a rod of arbitrary length and orientation intersects the vertical line can be calculated by integration over all its allowed lengths and angles.
\begin{multline}\label{eq:integrals}
 \int\limits_{-\pi/2}^{\pi/2}\int\limits_{0}^{l_m}  \frac{ l \cos \theta}{L} f_l(l) f_\theta (\theta) \,dl \,d\theta =\\ \frac{1}{L} \int\limits_{-\pi/2}^{\pi/2} f_\theta(\theta) \cos \theta \, d\theta \int\limits_{0}^{l_m}  l f_l(l) \, dl = \frac{\langle l \rangle \langle \cos \theta \rangle }{L}.
\end{multline}
Thus, the probability that a rod intersects a cell located on its right is
 $$
 P^\prime_2 = \frac{\langle l \rangle \langle \cos \theta \rangle }{m L}
 $$
Similarly, the probability that a rod intersects a cell located below it is
 $$
 P^{\prime\prime}_2 = \frac{\langle l \rangle \langle \cos \theta^\prime \rangle }{m L},
 $$
 where $\theta^\prime$ is the angle relative to the ordinate. Thus, the probability that a rod intersects a cell is
 $$P_2 = P^\prime_2 + P^{\prime\prime}_2 = \frac{\langle l \rangle(\langle \cos \theta \rangle +  \langle \cos \theta^\prime \rangle )}{m L}.$$
Finally, the fraction of empty cells is
\begin{equation}\label{eq:transm}
 \frac{M}{m^2} = \left( 1 - m^{-2} - \frac{\langle l \rangle(\langle \cos \theta \rangle +  \langle \cos \theta^\prime \rangle )}{m L}\right)^N.
 \end{equation}
 The quantity $T = M m^{-2}$ may be treated as the transmittance of the film. For isotropic deposition of rods, $\langle \cos \theta \rangle =  \langle \cos \theta^\prime \rangle = 2/\pi$, hence,
 \begin{equation*}\label{eq:transmiso}
 T = \left( 1 - m^{-2} - 4(\pi m L)^{-1}\right)^N.
 \end{equation*}
 For completely aligned rods (nematic order), $\langle \cos \theta \rangle = 1$, $ \langle \cos \theta^\prime \rangle = 0$, hence,
 \begin{equation*}\label{eq:transmalign}
 T = \left( 1 - m^{-2} - (m L)^{-1}\right)^N.
 \end{equation*}
  Let $L \to \infty$ and $m \to \infty$ in such a way that $k^\ast = m/L = \mathrm{const}$. Let $L = c \langle l \rangle$ then
 \begin{multline*}
 \lim_{c \to \infty} T = \\ \lim_{c \to \infty} \left[ 1 + \frac{1}{c^2} \left(- \frac{1}{(k^\ast \langle l \rangle)^2} - \frac{\langle \cos \theta \rangle +  \langle \cos \theta^\prime \rangle }{k^\ast  \langle l \rangle } \right) \right]^{n c^2 \langle l \rangle^2} =\\
 \exp\left\{ - \frac{n}{k^\ast}\left[ \frac{1}{k^\ast } + \langle l \rangle \left(\langle \cos \theta \rangle +  \langle \cos \theta^\prime \rangle\right) \right]  \right\}.
\end{multline*}

\begin{figure}[!htb]
  \centering
  \includegraphics[width=\columnwidth]{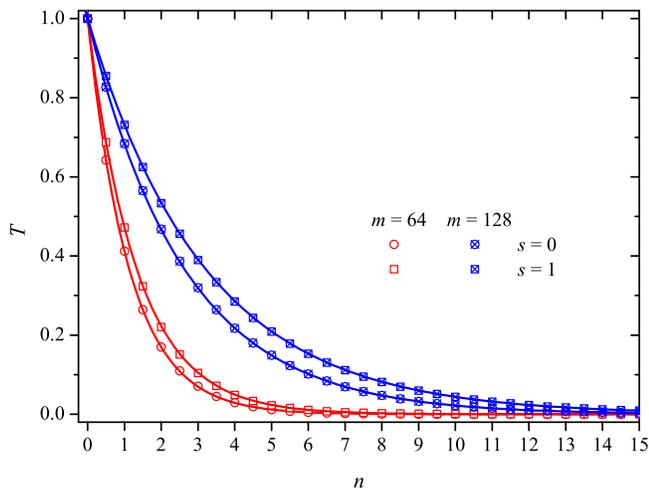}
  \caption{Examples of transmittances, $T$, vs the number density, $n$, for an isotropic angular distribution ($s = 0$) and for completely aligned rods ($s = 1$). The curves correspond to Eq.~\eqref{eq:transm}. The length distribution obeys Eq.~\eqref{eq:lognorm}, $\sigma(l) = 1$.\label{fig:Tvsndisp}}
\end{figure}

Figure~\ref{fig:Tvsndisp} demonstrates how the transmittance, $T$, depends upon the number density, $n$, for the two limiting cases, viz., for an isotropic angular distribution ($s = 0$) and for completely aligned rods ($s = 1$). In both cases, the length distribution obeys Eq.~\eqref{eq:lognorm} with $\sigma(l) = 1$. The curves correspond to Eq.~\eqref{eq:transm}.

\subsection{Electrical conductance in the limiting case of a dense RRN (Model~II)}

\begin{figure}[!htb]
  \centering
  \includegraphics[width=\columnwidth]{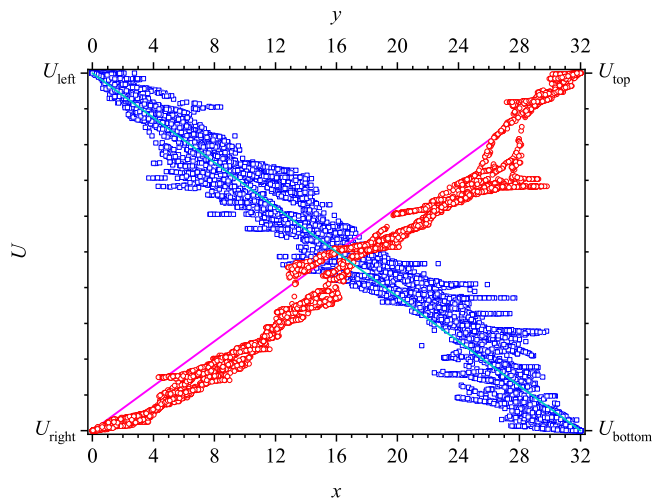}
  \caption{The potential of each node in the system is here plotted against its position in the sample (Model~II). $\sigma(l) = 1$, $s=0.8$. The rods are aligned along the abscissa. The potential difference is applied along the abscissa (left and bottom axes; horizontal positions of nodes are indicated) and along the ordinate (right and upper axes; vertical positions of nodes are indicated). \label{fig:voltage}}
\end{figure}

The results by Kumar et al.\cite{Kumar2016JAP,Kumar2017JAP} for the electrical conductance of dense 2D RRNs can be extended to anisotropic RRNs. The main idea is based on the assumption that, in a dense homogeneous RRN, the potential drop is linear.\cite{Kumar2016JAP,Kumar2017JAP} This assumption also holds in the case of anisotropic RRNs. Figure~\ref{fig:voltage} demonstrates an example of the potential distribution calculated using Model~II in a  system with $\sigma(l) = 1$ and $s=0.8$ when the rods are aligned along the horizontal direction. When a potential difference is applied between the left and right borders of the system, the potential of nodes changes almost linearly with the horizontal position of each junction; due to the length dispersity of the fillers, these points form a wide band.  When the potential difference is applied between the upper and bottom borders of the system, the potential of nodes changes again almost linearly with the position of each junction (vertical ones in this case). However, the points form a narrow band, since, in the direction perpendicular to the alignment of fillers, their length dispersity has insignificant effect on the potential distribution. Thus, in either direction, the potential drop is close to linear.

Let us postpone consideration of our 2D system with rodlike fillers for a while and consider a dense 2D RRN, assuming a graph of this RRN is identical to its backbone. Let the total number of edges of this graph be $N_E$. When the RRN is subject to a potential difference, $U$, applied along the horizontal axis [Fig.~\ref{fig:mesh}], the potential drop across an edge of length $l_E$ is $U l_E \cos \theta/L$. When the conductance per unit length of the edge is $\gamma_0$ then the electrical current through the edge is $\gamma_0 U l_E \cos \theta/ L$. Hence, by close analogy with~\eqref{eq:integrals}, the total current through the system is
\begin{multline*}
i =  N_E \int\limits_{-\pi/2}^{\pi/2}\int\limits_{0}^{ l_m} \frac{\gamma_0 U  \cos \theta}{ L } \frac{  l_E  \cos \theta}{L} f_l( l_E ) f_\theta (\theta) \,d l_E  \,d\theta =\\
 = n_E U \gamma_0  \int\limits_{-\pi/2}^{\pi/2} f_\theta(\theta) \cos^2 \theta \, d\theta \int\limits_{0}^{ l_m} l_E  f_l( l_E ) \, d l_E  =\\=  n_E U \gamma_0 \langle  l_E  \rangle \langle \cos^2 \theta \rangle,
 \end{multline*}
where $n_E = N_E/ L^2$ is the number of edges per unit area. Taking into account~\eqref{eq:s}, the conductance of the 2D RRN~is
\begin{equation}\label{eq:condanis}
G_i = \frac{ \gamma_0  n_E \langle l_E \rangle ( 1 \pm s ) }{2}, \quad i = \parallel, \perp.
\end{equation}
Returning to the original problem of the electrical conductance of the RRN produced by the deposition of rods, we need to know how $N_E$ and $\langle l_E \rangle$ depend on the number of deposited rods, $N$. Moreover, the fraction of edges belonging to the backbone is needed. If this information  is anyhow obtainable, the conductance can be calculated using~\eqref{eq:condanis}. For instance, in computer simulations, this information is always directly available.

\begin{figure}[!htb]
  \centering
  \includegraphics[width=\columnwidth]{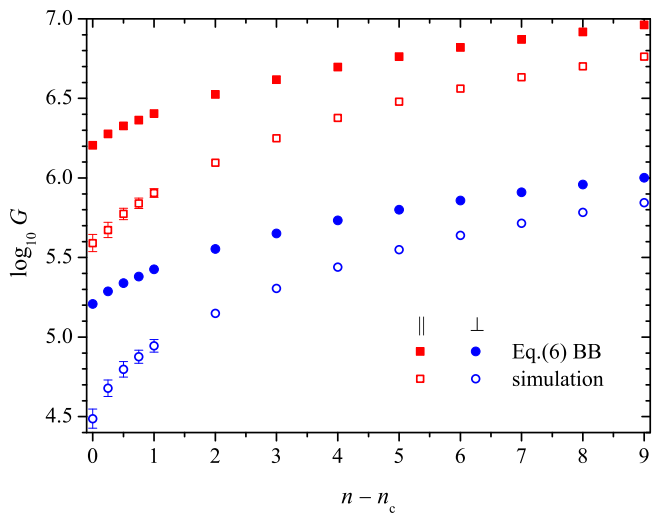}
  \caption{Comparison of the electrical conductance obtained from Model~II and calculated using Eq.~\eqref{eq:condanis}  for samples with $\sigma(l) = 1$, $s=0.8$.\label{fig:analytics}}
\end{figure}

Figure~\ref{fig:analytics} compares the electrical conductance obtained from Model~II and calculated using Eq.~\eqref{eq:condanis}  for samples with $\sigma(l) = 1$, $s=0.8$. The values of $n_E$ and $\langle l_E \rangle$ correspond to the backbone of the percolation cluster. Eq.~\eqref{eq:condanis} predicts overestimated values of the electrical conductance  compared to the simulations. This deviation reduces as the concentration of fillers increases. The presumptive origin of such deviation is because of the non strictly linear drop of the potential [see Fig.~\ref{fig:voltage}].

\subsection{Electrical conductance: Comparison of the two approaches}\label{subsec:comparison}
\begin{figure}[!htb]
  \centering
  \includegraphics[width=\columnwidth]{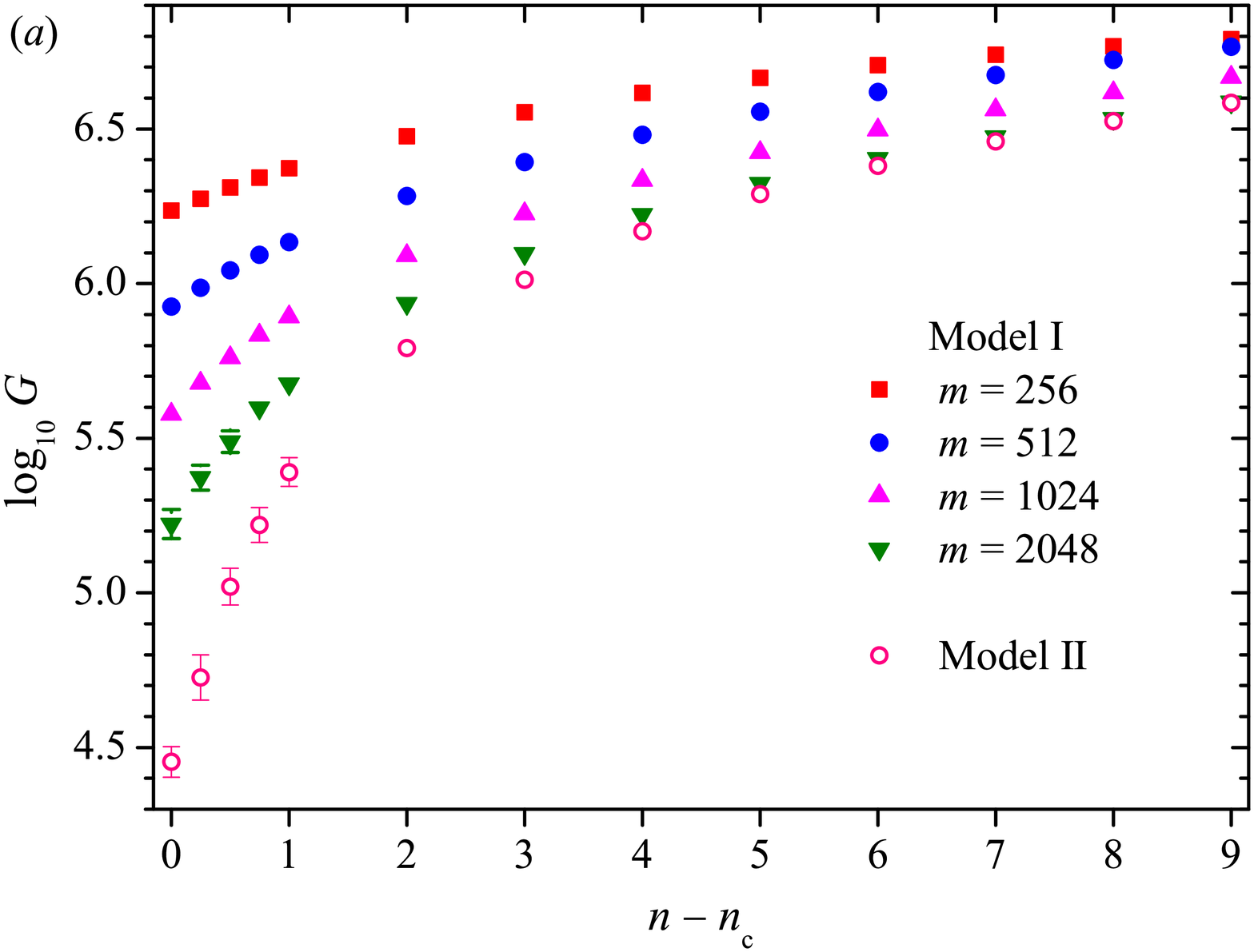}\\
    \includegraphics[width=\columnwidth]{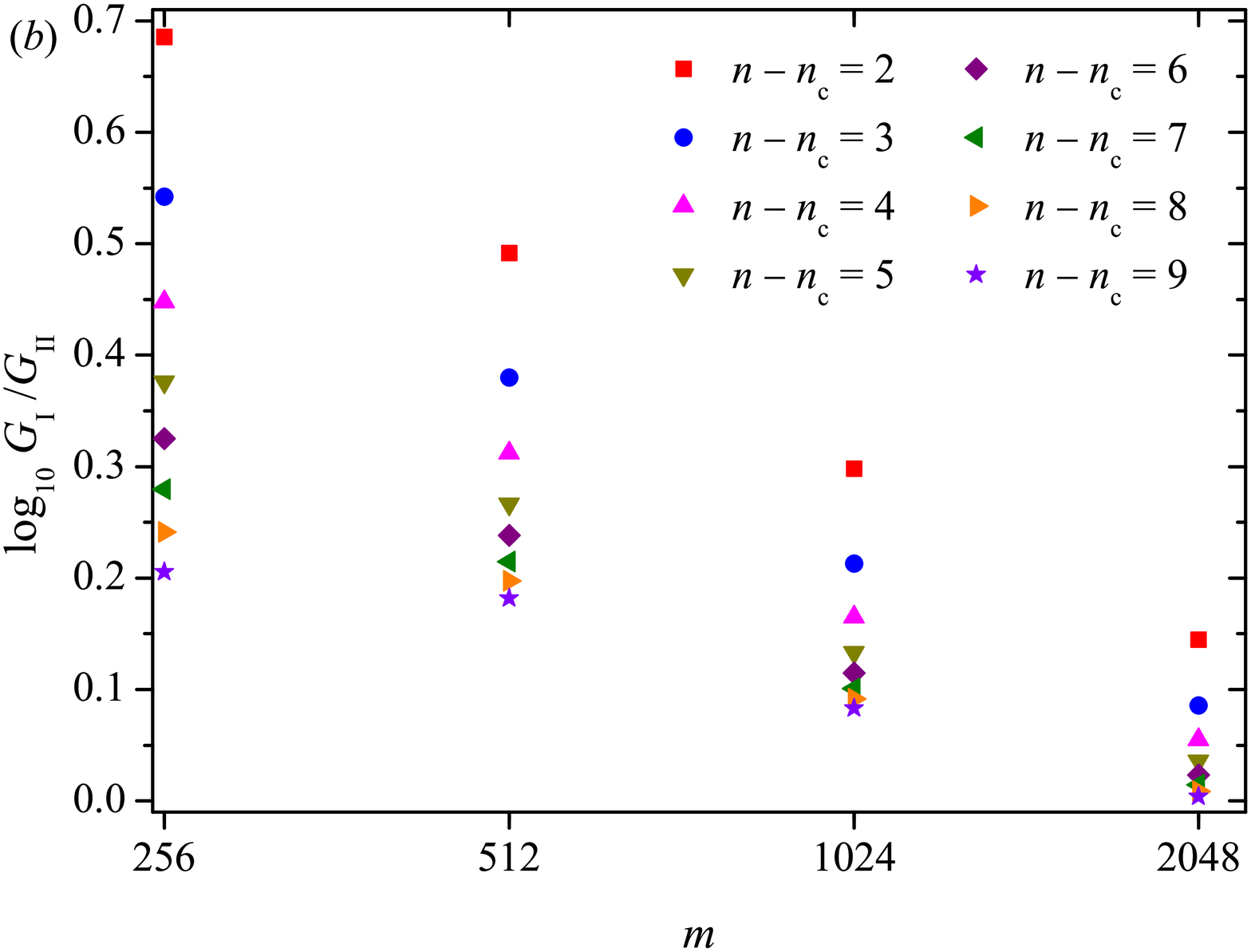}\\
  \caption{Comparison of the electrical conductance for different values of the density of the mesh, $m$, for isotropic system with rods of equal size, as calculated within Model~I, $G_\text{I}$, and within Model~II, $G_\text{II}$. The results were averaged over 10 independent runs.  The error bar is of the order of the marker size when not shown explicitly. (a) Example of the electrical conductance  vs the number density.
  (b) Example of the ratio $G_\text{I}/G_\text{II}$  vs the density of the mesh, $m$.
\label{fig:conductivityvsn}}
\end{figure}

Figure~\ref{fig:conductivityvsn}(a) demonstrates an example of the dependencies of the electrical conductances on the number density, as calculated within Model~I and Model~II. The electrical conductance calculated using Model~I tends clearly to the value obtained using Model~II when the value of $m$ increases. Slightly above the percolation threshold, the values of the electrical conductance vary significantly as the value of $m$ changes. For large values of the number density, this difference is less pronounced. Essentially, electrical conductance predictions from  Model~I are overestimated compared with those from Model~II. Nevertheless, with increasing value of $m$, the electrical conductance calculated within Model~I tends to the electrical conductance  calculated within Model~II [Fig.~\ref{fig:conductivityvsn}(b)].

\begin{figure}[!htb]
  \centering
  \includegraphics[width=\columnwidth]{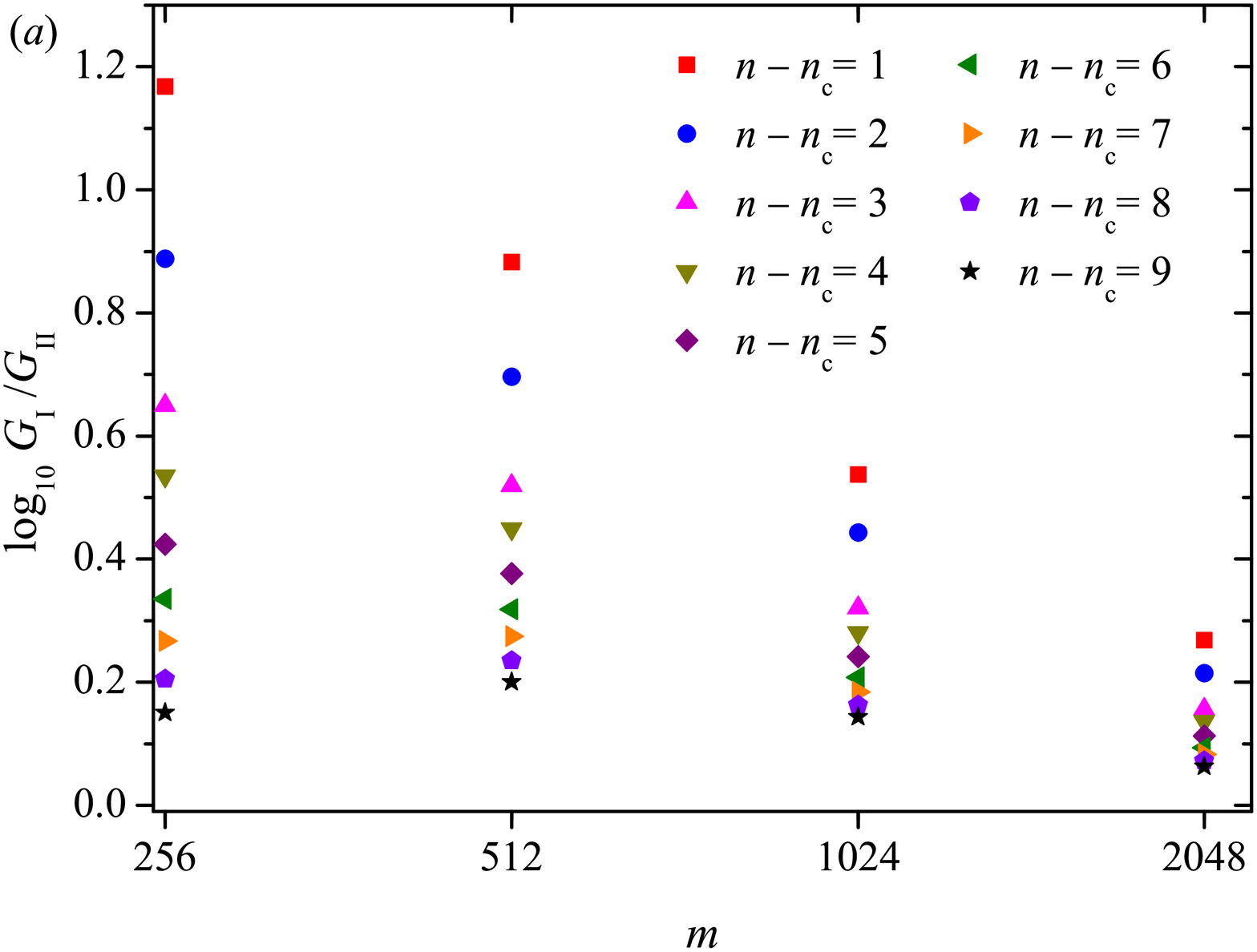}\\
  \includegraphics[width=\columnwidth]{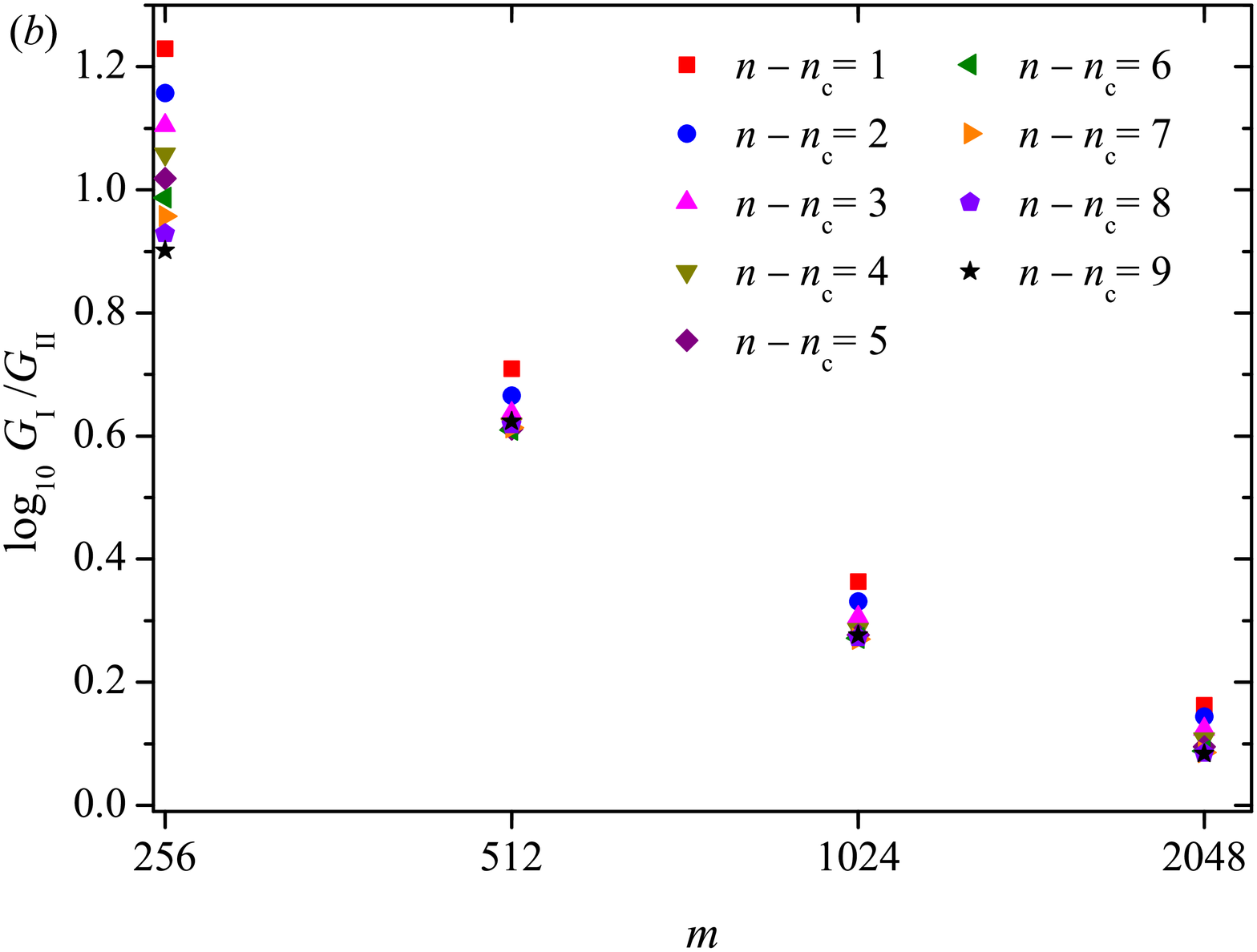}\\
  \caption{Example of the ratio of the electrical conductance $G_\text{I}$ calculated within Model~I to the electrical conductance $G_\text{II}$ calculated within Model~II vs the density of the mesh, $m$, for an  anisotropic system with length dispersity of the rods ($\sigma(l) = 1$, $s=0.8$). The results were averaged over 10 independent runs. (a) along the direction of rod alignment, (b) perpendicular to the direction of rod alignment. Error bars are of order of marker size.\label{fig:diffGvsm}}
\end{figure}

Similar behavior was observed in the case of aligned rods with length dispersity. Figure~\ref{fig:diffGvsm} demonstrates such an example for $s=0.8$ and $\sigma(l) = 1$.

\begin{figure}[!htb]
  \centering
  \includegraphics[width=\columnwidth]{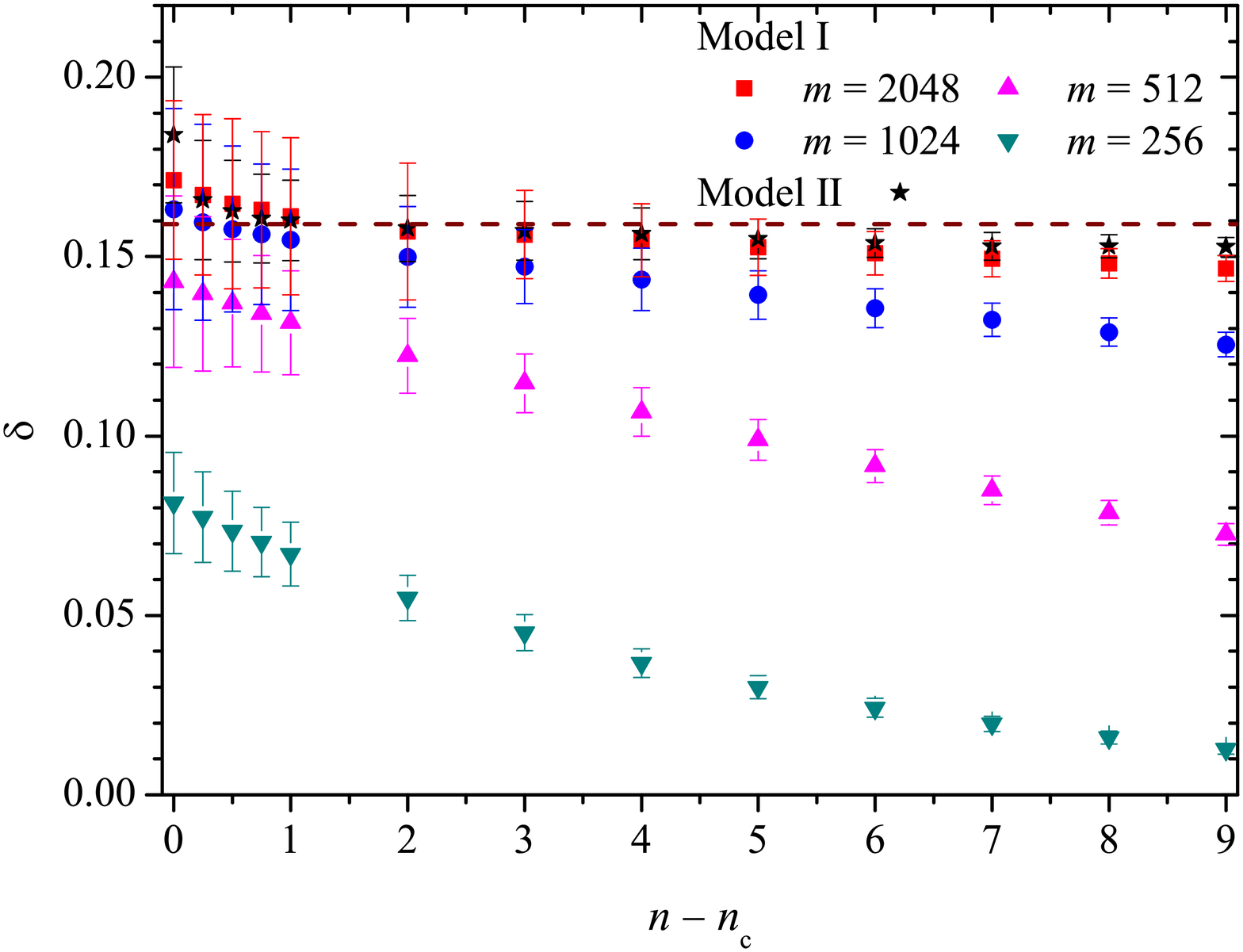}\\
    \caption{Example of the electrical anisotropy $\delta$ calculated within Model~I and within Model~II for different values of the density of the mesh, $m$, vs the number density for an anisotropic system with length dispersity of the rods ($\sigma(l) = 1$, $s=0.8$). The results were averaged over 10 independent runs. Error bars are of the order of the marker size. The dashed line corresponds to the estimation obtained using Eq.~\eqref{eq:deltaanal} $\delta \approx 0.159$.
    \label{fig:delta}}
\end{figure}

Figure~\ref{fig:delta} compares the electrical anisotropy calculated within Models~I and II. The electrical anisotropy is more pronounced in the vicinity of the percolation threshold, then, it decreases as the number density decreases.  Far above the percolation threshold ($n \gg n_c$), the results obtained within Model~I clearly tend to the results obtained within Model~II as the density of the mesh increases. The values of the electrical anisotropy calculated within both models clearly tend to the value predicted by Eq.~\eqref{eq:condanis}, i.e.
\begin{equation}\label{eq:deltaanal}
\delta = \frac{\log_{10} \left[(1 + s ) / ( 1  - s)\right]}{\log_{10} \Delta}.
\end{equation}
Figure~\ref{fig:delta} suggests that the system under consideration with $n \approx 10 n_c$ can be treated as a dense RRN.

\section{Conclusion}\label{sec:concl}
We considered 2D systems composed of randomly distributed, aligned, conductive rods with length dispersity. We found an analytical relationship between the number density of the fillers and the transparency of the system [Eq.~\eqref{eq:transm}]. The analytical relationship for the electrical conductance of isotropic dense RRNs~\cite{Kumar2016JAP} has therefore been extended to cover anisotropic, dense RRNs identical to their backbones [Eq~\eqref{eq:condanis}].

The electrical conductances calculated within two approaches have been compared. We found that, above the percolation threshold, the conductances exhibit similar behavior in both approaches. However, comparison of the two models suggests that, when the conductance of the host matrix has to be taken into consideration, i.e., when Model~I is preferable, a fairly fine mesh should be applied to obtain a reasonable estimation of the electrical conductance. Nevertheless, overestimation of the electrical conductance should be kept in mind.

We found that, in anisotropic system, the length dispersity of rods has a more pronounced effect on the electrical properties along the direction of rod alignment.

\begin{acknowledgments}
We acknowledge the funding from the Ministry of Science and Higher Education of the Russian Federation, Project No.~3.959.2017/4.6.
\end{acknowledgments}

\bibliography{dispersity}

\end{document}